\documentclass[useAMS,usenatbib]{mn2e}
\usepackage{graphicx}
\title[Trends with Condensation Temperature]{Parent Stars of Extrasolar Planets. XI. Trends with Condensation Temperature Revisited}
\author[G.\ Gonzalez et al.]{G.\ Gonzalez$^{1}$, M. K.\ Carlson$^{1}$, R. W.\ Tobin$^{2}$\\
$^{1}$Grove City College, Rockwell Hall, 100 Campus Drive, Grove City, PA 16127, USA\\
$^{2}$Department of Physics and Astronomy, Ball State University, Muncie, IN 47306 USA\\
}

\begin{document}

\date{Accepted ??. Received ??; in original form ??}

\pagerange{\pageref{firstpage}--\pageref{lastpage}} \pubyear{??}

\maketitle

\label{firstpage}

\begin{abstract}
We report the results of abundance analyses of new samples of stars with planets and stars without detected planets. We employ these data to compare abundance-condensation temperature trends in both samples. We find that stars with planets have more negative trends. In addition, the more metal-rich stars with planets display the most negative trends. These results confirm and extend the findings of Ramirez et al. (2009) and Melendez et al. (2009), who restricted their studies to solar analogs. We also show that the differences between the solar photospheric and CI meteoritic abundances correlate with condensation temperature.
\end{abstract}

\section{Introduction}

\citet{mel09} were the first to detect a significant correlation between elemental abundances and condensation temperature (T$_{\rm c}$) in a sample Sun-like stars; in a follow-up study, \citet{ram09} confirmed their findings. In particular, they found solar twins/analogs to be enhanced in refractory elements relative to the Sun. Although this was the first time such a trend had been found, it had been searched for unsuccessfully multiple times before within the context of the self-enrichment hypothesis \citep{hu05, gg06, ecu06}. \citet{ram09} speculate that the trends are due to planet formation processes.

In order to test the findings of \citet{ram09} and also to expand on their analysis over a broader range in T$_{\rm eff}$, we revisit the topic of abundance trends with T$_{\rm c}$ among stars with planets (SWPs). We do so with a new method of analysis we introduced in \citet{gg08} and new stellar samples, which we described in \citet{gg10}.

Our paper is organized as follows. In Section 2 we describe the new SWP and comparison star samples. We employ them in Section 3 to search for evidence of abundance trends with T$_{\rm c}$; we also examine the recent abundance data of \citet{neves09} and the Solar System abundances. We discuss our results in Section 4 and present our conclusions in Section 5.

\section{Description of Data}

In \citet{gg10} we described our most recent SWP and comparison stars samples and presented the results of our new fine abundance analyses. We compared the Li abundances of the SWPs to comparison stars using the method of comparison introduced in \citet{gg08} and confirmed that SWPs near the solar temperature have lower Li abundances than similar stars not known to harbor Doppler detectable planets. We deferred reporting the results of our analyses of other elements in these samples until the present paper. For detailed descriptions of the observations, data reduction and abundance analyses, the reader is referred to \citet{gg10}. 

Our initial SWP and comparison stars samples contain 85 and 59 stars, respectively. We list their abundances in Table 1 (provided online). In order to have the best chance of detecting the subtle trends between abundance (as [X/H]) and T$_{\rm c}$, we only calculated the abundances of those elements represented by 2 or more quality absorption lines in our spectra: Na (2), Al (2), Si (6), Ca (2), Sc (2), Ti (5), V (6), Cr (3), Fe (53), Co (3), Ni (6). We employ these data in our analyses below.

\section{Searching for Trends with T$_{\rm c}$}

\subsection{New Data}

In \citet{gg08} we introduced a new index, $\Delta_1$, which is a measure of the distance between two stars in T$_{\rm eff}$-[Fe/H]-log g-M$_{\rm v}$ space. In that study we calculated a weighted average Li abundance difference between a given SWP and all the stars in the comparison sample, with $(\Delta_1)^{-2}$ serving as the weight. We also applied this method (with the addition of corrections for bias) in \citet{gg10}. We employ essentially the same analysis method here, only that we are comparing [X/H]-T$_{\rm c}$ slopes rather than Li abundances.

The T$_{\rm c}$ values are from \citet{lod09}. The elements we measured span the T$_{\rm c}$ range 958 K (Na) to 1659 (Ca, Sc). This is about the same range of T$_{\rm c}$ that \citet{ram09} reported finding an [X/H]-T$_{\rm c}$ trend.

We further restricted the samples in the present analysis by eliminating stars that do not satisfy the following criteria: $5400 <$ T$_{\rm eff} < 6300$ K, uncertainty in T$_{\rm eff} < 70$ K, available {\it Hipparcos} parallax. After this culling, the SWP and comparison stars samples contain 65 and 56 stars, respectively.

For each star in our two samples, we calculated the slope between abundance (as [X/H]) and T$_{\rm c}$ using standard linear least-squares; the slope values are listed in the online table. The average uncertainty in the slope is about $\pm 0.00007$ dex K$^{\rm -1}$ and is the same for the SWPs and the comparison stars. We plot the weighted average [X/H]-T$_{\rm c}$ slope differences between SWPs and comparison stars in Figure 1. Overall, SWPs appear to display more negative [X/H]-T$_{\rm c}$ slopes than the comparison stars, the primary exception being some SWPs near the solar temperature.

\begin{figure}
  \includegraphics[width=3.5in]{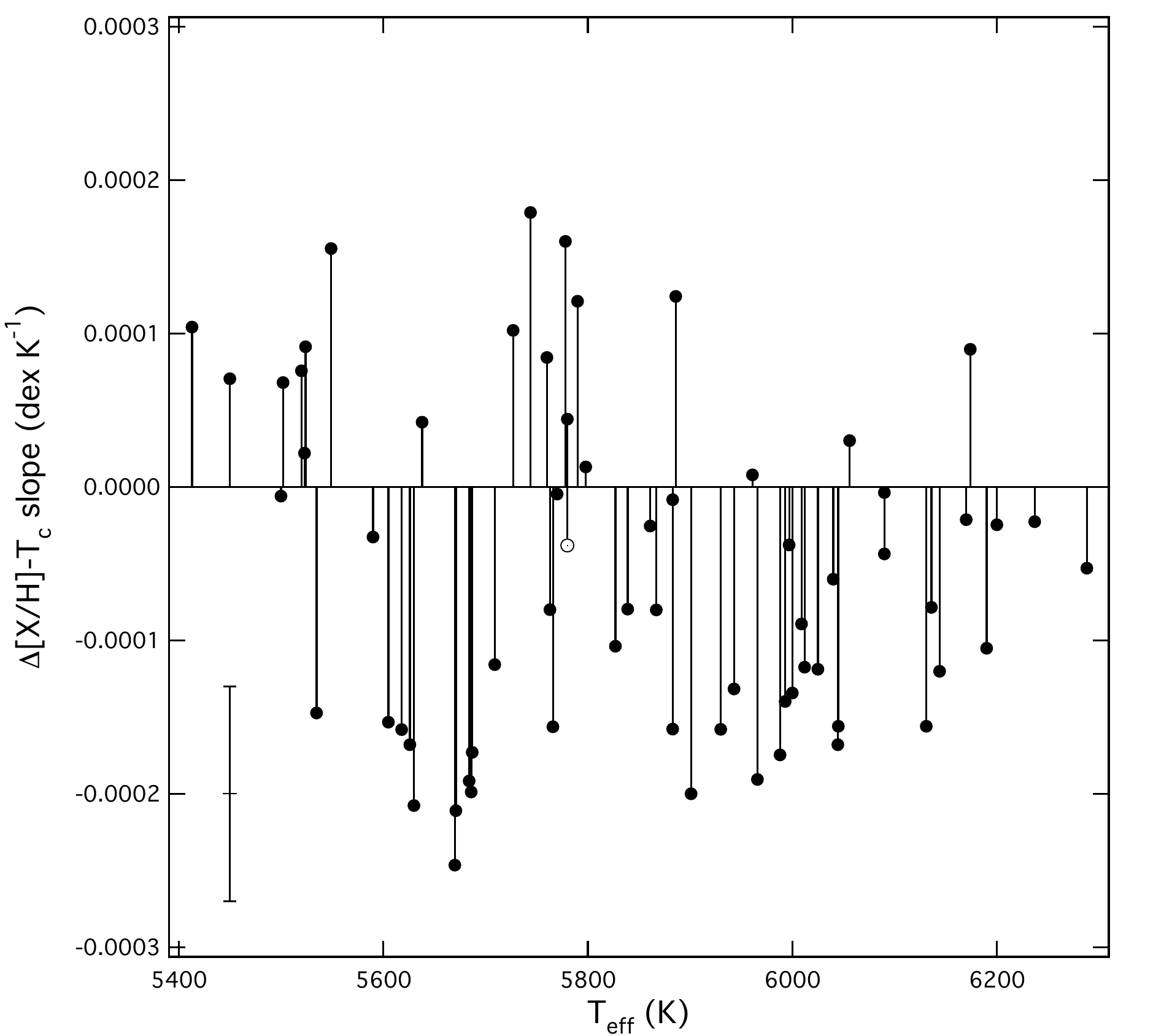}
 \caption{Weighted-average [X/H]-T$_{\rm c}$ slope differences between SWPs and comparison stars using the samples in the present study. The open circle with the dot represents the solar value. The error bars for the typical SWP slope difference are shown in the lower left.}
\end{figure}

When interpreting the datum corresponding to the Sun in each of Figure 1 and Figure 2 (and additional figures in the following sections), it is important to keep the following points in mind. First, although the [X/H]-T$_{\rm c}$ slope for the Sun is identically zero by definition, the weighted {\it difference} slope does not fall on 0.0 in the figures. The fact that it is negative tells us that the Sun's slope is more negative than the slopes of similar comparison stars. Second, the uncertainty in the [X/H]-T$_{\rm c}$ slope for the Sun is also zero, but the difference slope plotted in the figures will have some uncertainty associated with it due to uncertainties in the slopes of the comparison stars. The effective uncertainty of the Sun's difference slope should be considerably less than $\pm 0.00007$ dex K$^{\rm -1}$. This implies its difference slope is significantly different from zero.

In \citet{gg10} we introduced an additional step in the analysis method of \citet{gg08} to correct a systematic bias inherent in the method. We have applied the same correction procedure to the present data and present the final corrected results in Figure 2 (see \citet{gg10} for a detailed description of the correction method). The basic trends evident in Figure 1 are not changed. 

The average SWP [X/H]-T$_{\rm c}$ slope value from the data in Figure 2 is $(-5.6\times10^{\rm -5}$ $\pm$ 11 (s.d.) 1.4 (s.e.m.))$\times10^{\rm -5}$ dex K$^{\rm -1}$. A simple count yields 46 SWPs with negative slopes and 19 with positive slopes. If we assume that slope difference values between -0.00007 and +0.00007 dex K$^{\rm -1}$ are not significantly different from zero and count only the values below and above this range, then there are 33 SWPs with negative slopes and only 8 with positive slopes.

\begin{figure}
  \includegraphics[width=3.5in]{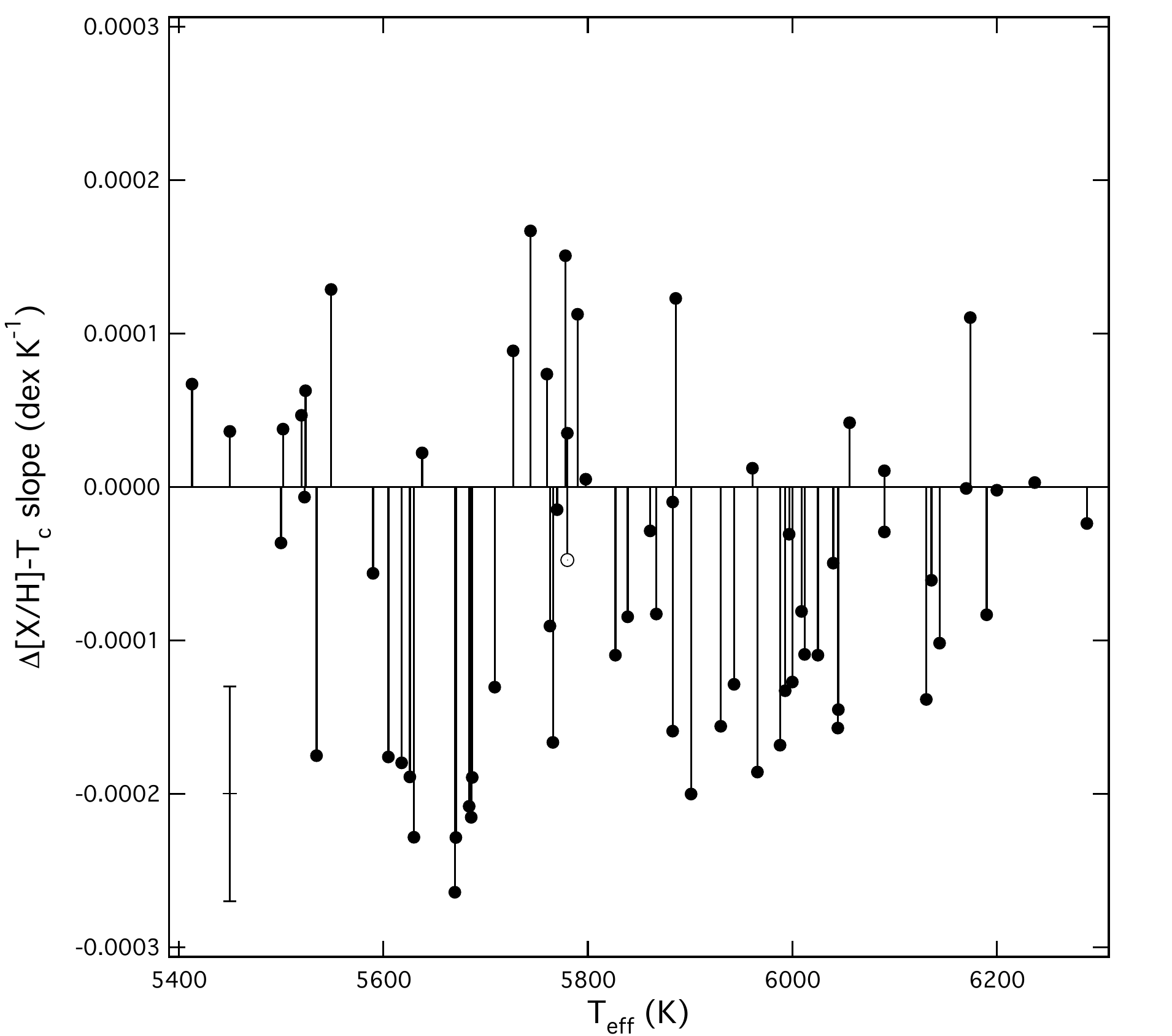}
 \caption{Same as Figure 1 but corrected for bias. See text for details.}
\end{figure}

Several stars in our samples can be considered solar analogs, which we define as having T$_{\rm eff}$, log g, [Fe/H], and M$_{\rm V}$ values within $\pm$ 100 K, 0.2 dex, 0.1 dex, and 0.2 magnitude, respectively, of the solar values. Two SWPs (HIP 15527, 80902) and 9 comparison stars satisfy our criteria. We plot the average relative abundances, as [X/Fe], against T$_{\rm c}$ for the SWP and comparison solar analogs in Figure 3.

\begin{figure}
  \includegraphics[width=3.5in]{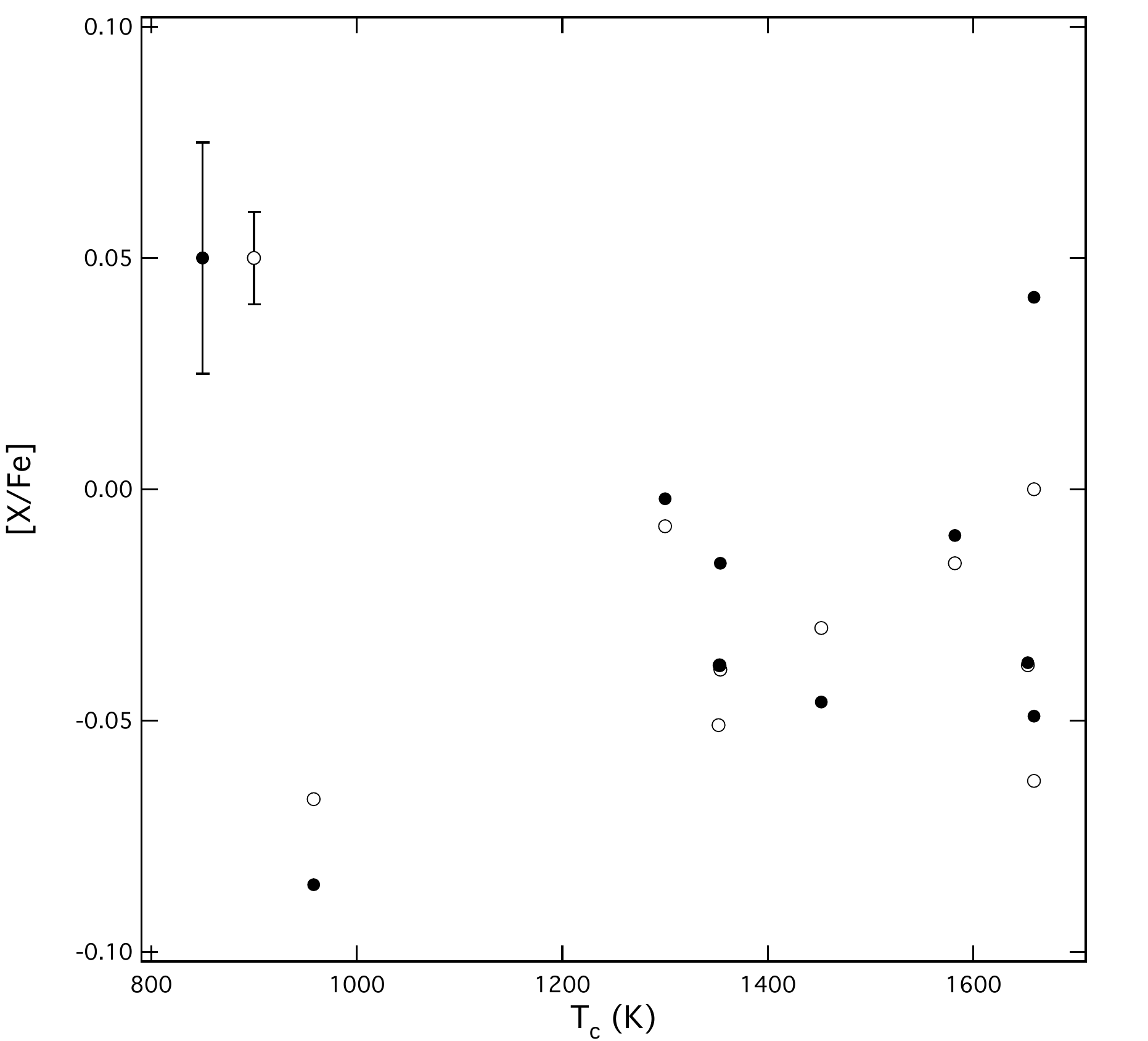}
 \caption{Average [X/Fe] values for the SWP (dots) and comparison (open circles) solar analog stars from our samples. The average error bars show the standard error of the mean for the SWPs and comparison stars; they were calculated from the star-to-star scatter (upper left).}
\end{figure}

The slopes of least-squares fits to the trends in Figure 3 are $(4.0 \pm 3.3) \times 10^{\rm -5}$ and $(9.6 \pm 6.1) \times 10^{\rm -5}$ dex K$^{\rm -1}$ for the comparison and SWP solar analogs, respectively. This shows that SWPs have an average slope 2.4 times that of the comparison stars. This is very close to the ratio of 2.2 found by \citet{mel09}. While our slope values cannot be directly compared to theirs (they included a larger range in T$_{\rm c}$), it should be safe to compare the ratios of slopes.

\subsection{\citet{neves09} Data}

\citet{neves09} published abundance data for 451 stars comparable in quality to ours. They measured the same elements we did (with the addition of Mg and Mn)\footnotetext{Note, we did not include the \citet{neves09} abundance values for Mn in our analysis, as they tend to have larger uncertainties than the other elements they measured.}. We applied the same culling criteria and analysis method as described above to their data, resulting in subsamples with 53 SWPs and 225 comparison stars. We show the bias-corrected weighted-average abundance-T$_{\rm c}$ slope differences in Figure 4.

\begin{figure}
  \includegraphics[width=3.5in]{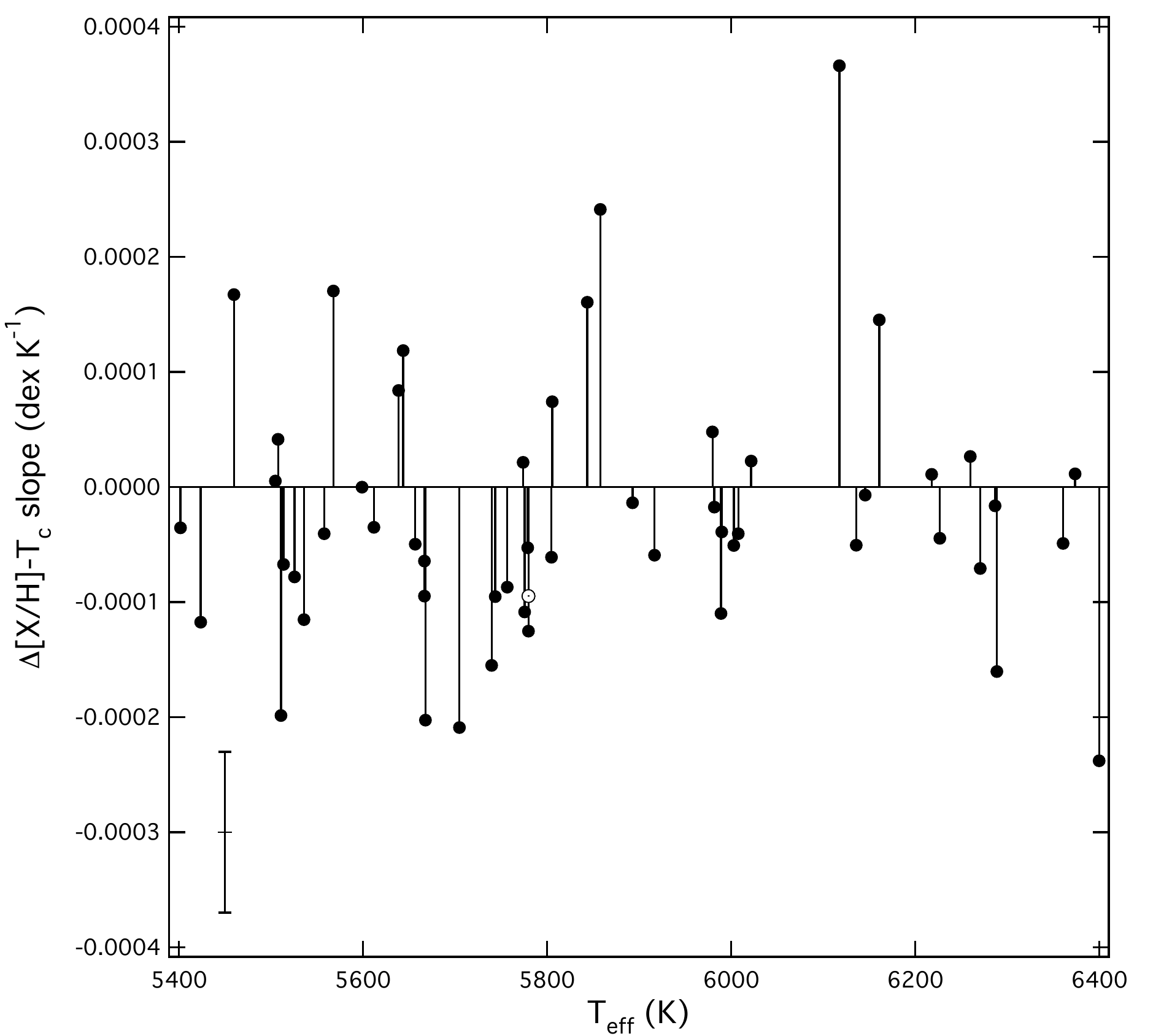}
 \caption{Same as Figure 2 but using the abundance data from the \citet{neves09} subsamples.}
\end{figure}

While there are fewer SWPs plotted in Figure 4 compared to Figure 2, similar negative average slopes are evident in the figures. The average SWP [X/H]-T$_{\rm c}$ slope difference value from these data is $-2.3\times10^{\rm -5}$ $\pm$ 11 (s.d.) 1.5 (s.e.m.)$\times10^{\rm -5}$ dex K$^{\rm -1}$; 36 SWPs have negative slopes, and 18 have positive slopes.

Three SWPs (HIP 15527, 80337, 116906) and 14 comparison stars from \citet{neves09} qualify as solar analogs according to our criteria. We plot the average relative abundances, as [X/Fe], against T$_{\rm c}$ for the SWP and comparison solar analog stars in Figure 5.

\begin{figure}
  \includegraphics[width=3.5in]{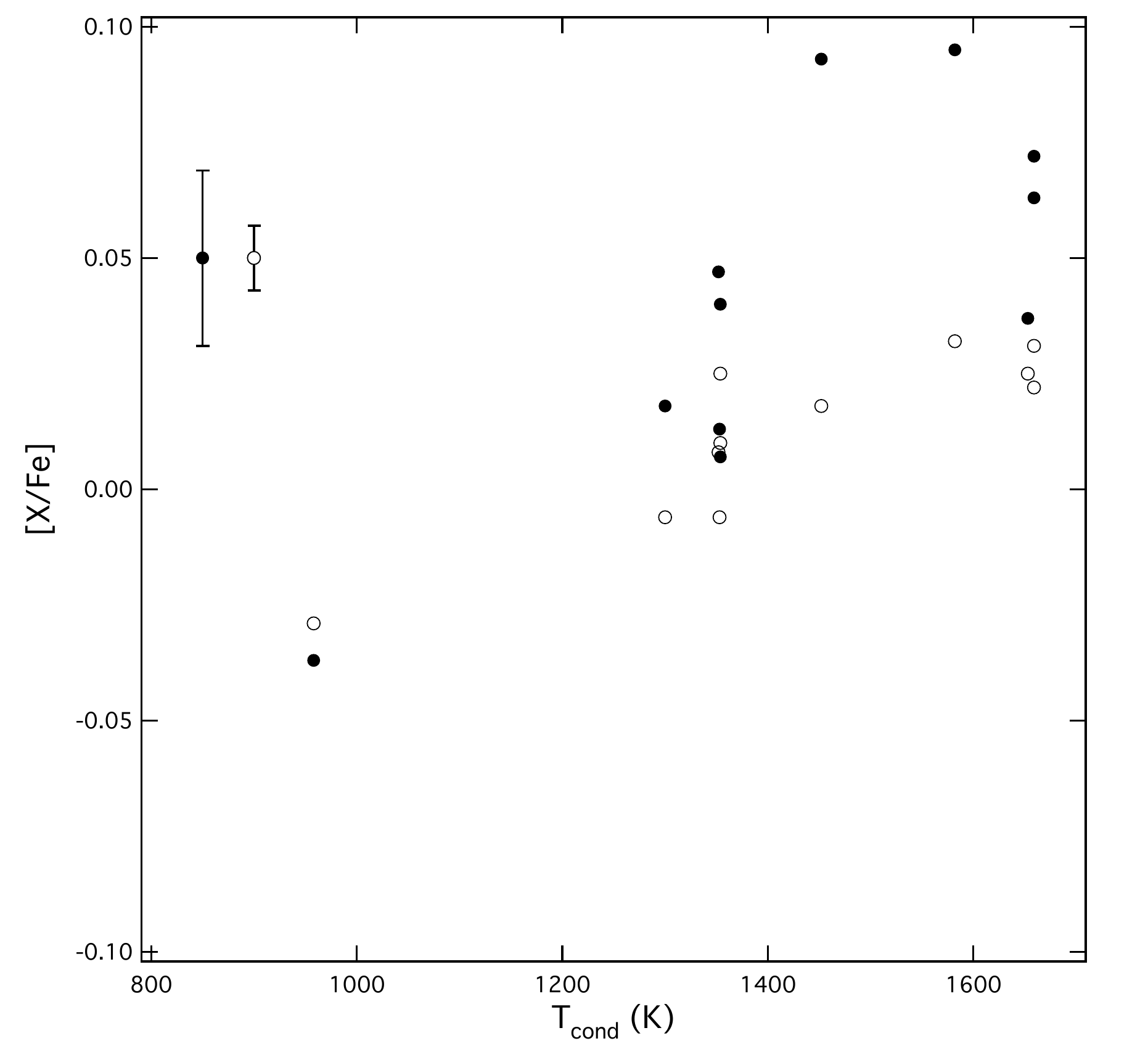}
 \caption{Same as Figure 3 but using the abundance data from the \citet{neves09} solar analogs.}
\end{figure}

The slopes of least-squares fits to the trends in Figure 5 are $(8.0 \pm 1.4) \times 10^{\rm -5}$ and $(15 \pm 4) \times 10^{\rm -5}$ dex K$^{\rm -1}$ for the comparison and SWP solar analogs, respectively. This gives a ratio of 1.9 for SWPs relative to comparison stars, again consistent with the results of \citet{mel09}.

\subsection{Combined Data}

There are 22 stars in common between \citet{neves09} and the present work. We find good agreement for the T$_{\rm eff}$ and abundance-T$_{\rm c}$ slope values for these stars. The mean difference in T$_{\rm eff}$ is $-16 \pm 40$ K. We compare the slopes in Figure 6. 

\begin{figure}
  \includegraphics[width=3.5in]{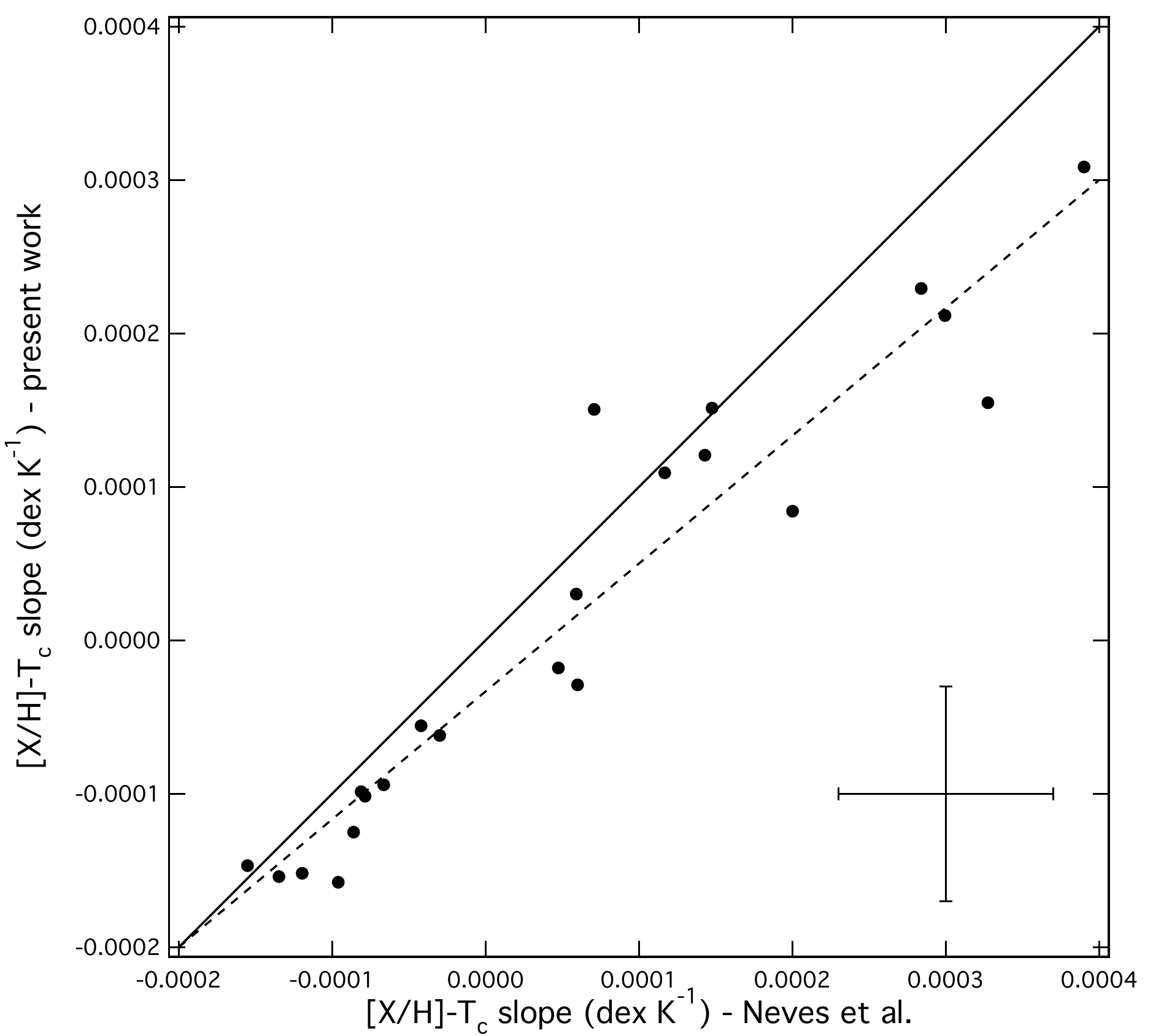}
 \caption{Comparison of the [X/H]-T$_{\rm c}$ slope values calculated from the abundances in the present work and from those of \citet{neves09} for stars in common. The solid line represents a one-to-one correlation; the dashed line is a least-squares fit. Typical error bars are shown in the lower right.}
\end{figure}

We adjusted our T$_{\rm eff}$ and [X/H]-T$_{\rm c}$ slope values to be on the same scale as the \citet{neves09} data based on the above comparisons. We then averaged the data for the stars in common and formed new combined SWP and comparison stars samples, which contain 100 and 277 stars, respectively. The bias-corrected combined [X/H]-T$_{\rm c}$ slope difference values are plotted in Figure 7.

\begin{figure}
  \includegraphics[width=3.5in]{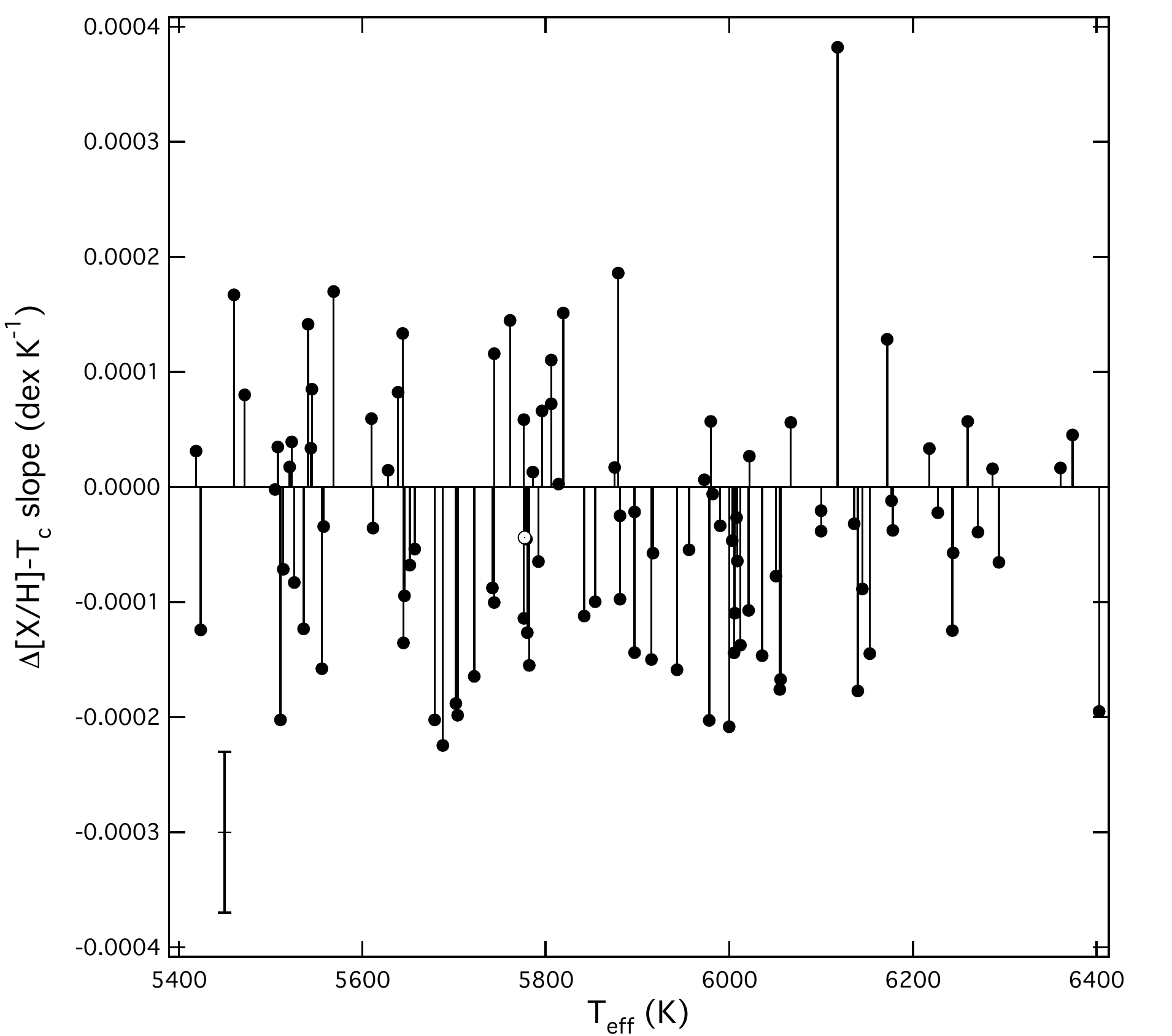}
 \caption{Same as Figure 2 but using the combined samples.}
\end{figure}

The average SWP abundance-T$_{\rm c}$ slope value from these data is $-4.0\times10^{\rm -5}$ $\pm$ 11 (s.d.) 1.1 (s.e.m.)$\times10^{\rm -5}$ dex K$^{\rm -1}$; 67 SWPs have negative slopes, and 34 have positive slopes. The number of stars with positive and negative slope values are comparable below about 5800 K, but the negative slopes strongly dominate above 5800 K. The Sun's slope difference is negative ($-4.0\times10^{\rm -5}$ dex K$^{\rm -1}$), but stars with both positive and negative slopes are present near the solar temperature.

\subsection{Solar System Abundances}

\citet{gg97} first noted a trend between the difference in the Solar System photospheric and meteoritic abundances and T$_{\rm c}$. \citet{gg06} revisited this anomaly using the abundance data of \citet{lod03} and \citet{asp05} and found that both data sets display a significant trend. Since 2006, new photospheric and meteoritic abundance data have been published. We repeat the analysis of \citet{gg06} below with more recent data to determine if the trend persists.

We show in Figure 8 the difference between the solar photospheric and meteoritic abundances plotted against T$_{\rm c}$ using the data from \citet{asp09}. The figure includes abundance differences with total uncertainties less than 0.10 dex. The slope from a weighted least-squares solution is $(7.3 \pm 2.5) \times10^{\rm -5}$ dex K$^{\rm -1}$, and the Pearson's correlation coefficient is 0.43. The probability that the slope is actually zero is about 1\%.

\begin{figure}
  \includegraphics[width=3.3in]{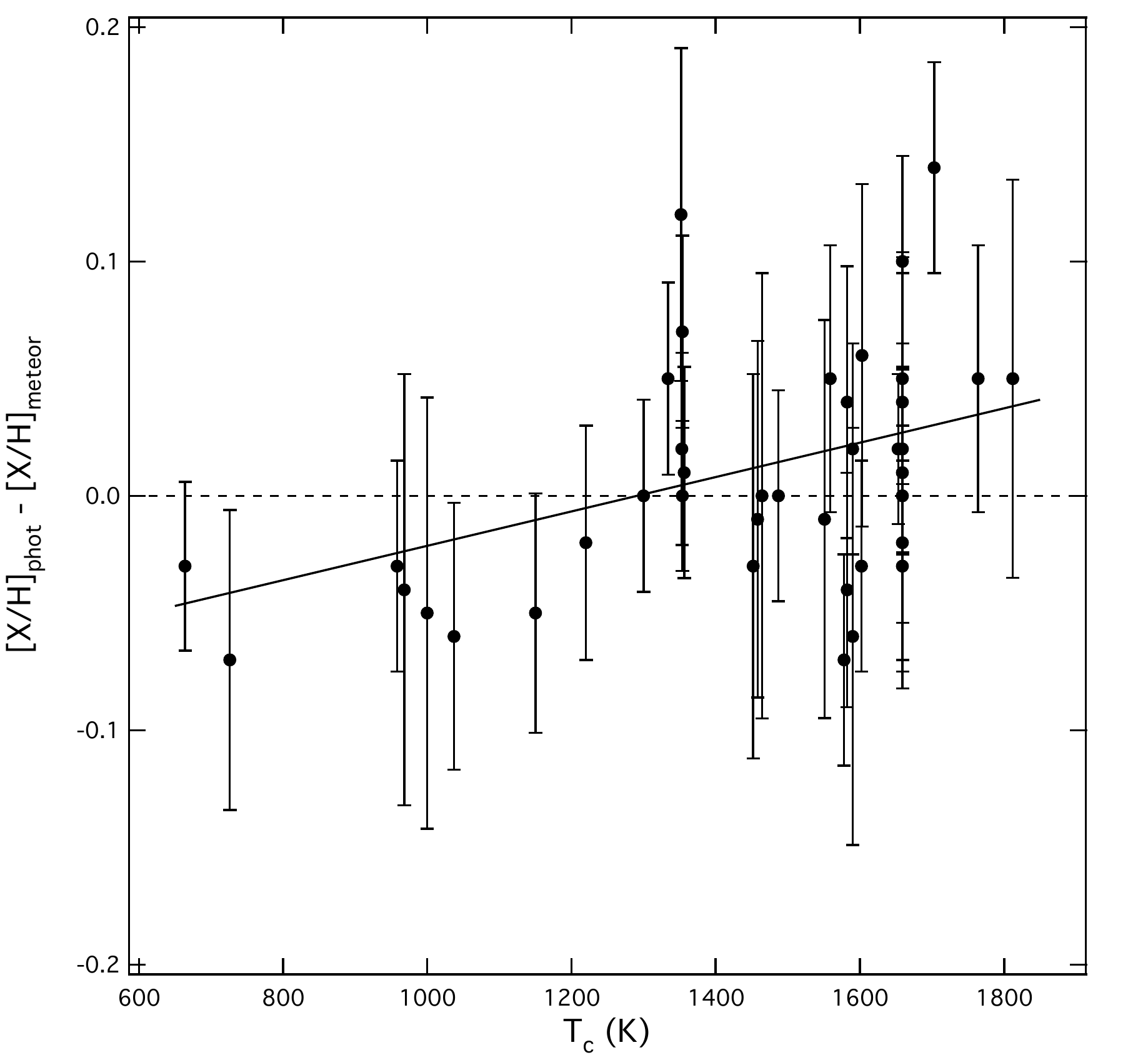}
 \caption{Difference between the solar photospheric and meteoritic elemental abundances is plotted against condensation temperature using the abundance data from Table 1 of \citet{asp09}. The error bars include the errors from both the solar and meteoritic abundances. The solid line is a weighted least-squares fit.}
\end{figure}

We plot in Figure 9 the difference between the solar photospheric and meteoritic abundances using the data compiled by \citet{lod09}.\footnotetext{The Co abundance we adopt in our analysis is from \citet{berg09}, which was published too late for \citet{lod09} to include it in their compilation.} The best-fit slope to the abundance differences with total uncertainties less than 0.10 dex is $(4.2 \pm 1.8) \times10^{\rm -5}$ dex K$^{\rm -1}$, and the Pearson's correlation coefficient is 0.37. The probability that the slope is actually zero is about 2\%.

There is reason to suspect that the data plotted in Figure 9 underestimate the trend with T$_{\rm c}$. The data compiled by \citet{lod09} is a heterogeneous collection of recent abundance determinations from the literature. When multiple solar photospheric abundance determinations are available for a given element, \citet{lod09} often select the values that are in closest agreement with the meteoritic value. For example, when discussing one of the published the Na photospheric abundances, \citet{lod09} write, ''We do not adopt this value, as it is 25\% lower than the meteoritic value as well as previously determined photospheric Na abundances.'' They make similar statements for K and Os. On the other hand, for P, they note that the new abundance value agrees well with the meteoritic value. This approach necessarily biases the abundances towards better agreement with the meteoritic values.

If one takes a longer historical view of the study of Solar System abundances, it is true that as the uncertainties have been reduced for many elements, the photospheric and meteoritic abundance values have come to agree much more closely. It is a mistake, however, to assume that the photospheric and meteoritic abundances will continue to agree as the abundance data continue to improve.

\begin{figure}
  \includegraphics[width=3.3in]{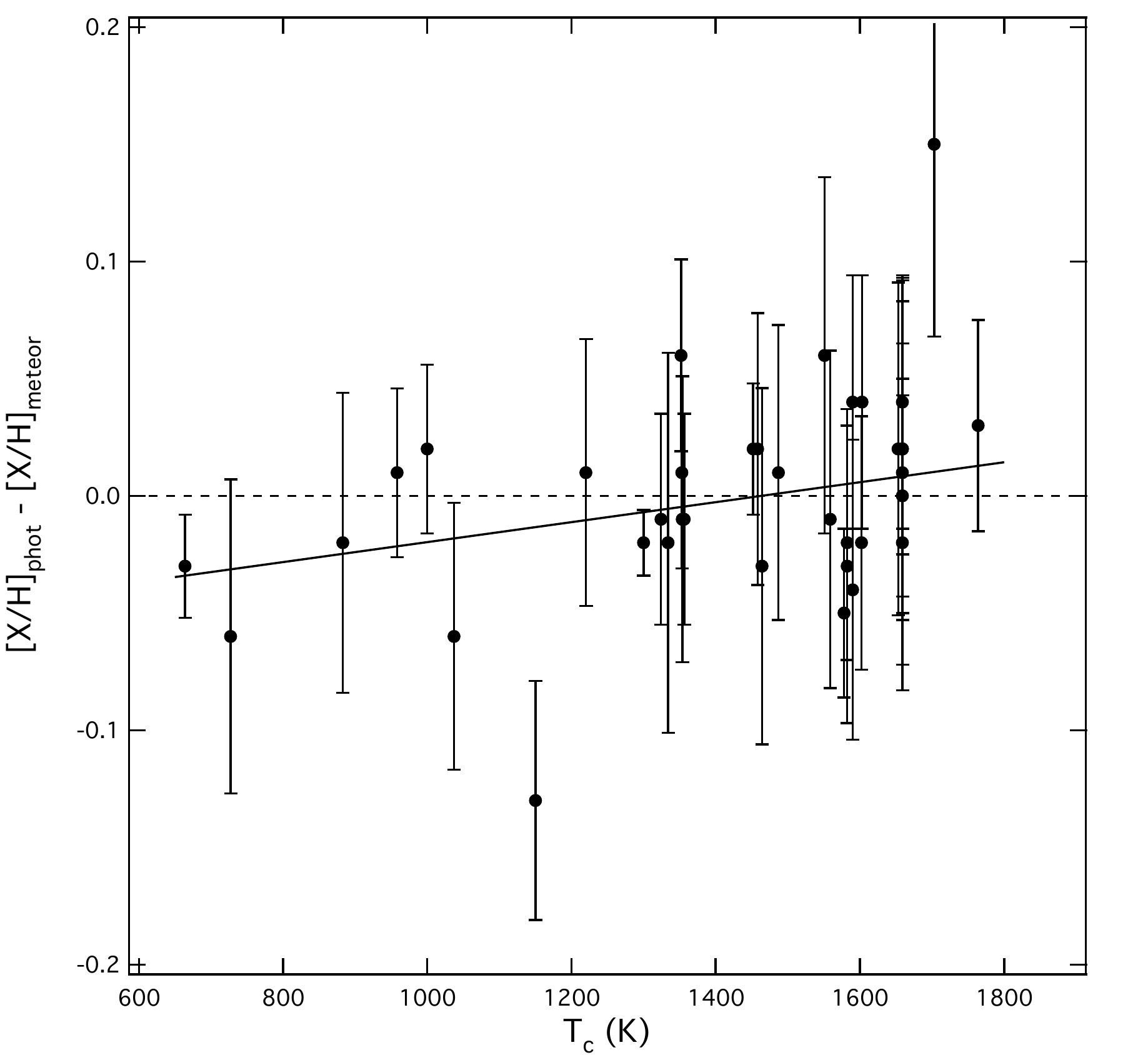}
 \caption{Same as Figure 6, but using the data from Table 4 of \citet{lod09}.}
\end{figure}

\citet{gg06} obtained slope values of $(8.2 \pm 2.7) \times10^{\rm -5}$ dex K$^{\rm -1}$ and $(6.1 \pm 2.5) \times10^{\rm -5}$ dex K$^{\rm -1}$ from the \citet{asp05} and \citet{lod03} data, respectively. 

As a kind of control, we repeated the analyses shown in Figures 8 and 9 with atomic number, Z,  replacing T$_{\rm c}$. The resulting slopes are $(-3 \pm 300) \times10^{\rm -6}$ dex Z$^{\rm -1}$ and $(3 \pm 3) \times10^{\rm -4}$ dex Z$^{\rm -1}$ from the \citet{asp09} and \citet{lod09} data, respectively. Neither of these trends is statistically significant. This gives us additional confidence that the trends in Figures 8 and 9 are real.

Finally, we examine the solar abundances determined by \citet{caf10}. They measured only 12 elements, including Li, C, N, O. Excluding these four elements and another element with a large uncertainty (Os), leaves us with the following seven elements: P, S, K, Fe, Eu, Hf and Th. A weighted least-squares fit gives a slope of $(8.5 \pm 6.6) \times10^{\rm -5}$ dex K$^{\rm -1}$. 

\section{Discussion}

Our abundance data and the abundance data of \citet{neves09} separately show that SWPs have more negative [X/H]-T$_{\rm c}$ slopes than stars without known planets; the combined data display similar patterns. These results imply that SWPs are relatively more depleted in high-T$_{\rm c}$ elements than low-T$_{\rm c}$ elements relative to comparison stars.

\citet{mel09} were the first to detect significant correlations between abundances and T$_{\rm c}$ among Sun-like stars. Restricting their study to solar analogs and twins, they found that the stars in their sample (with and without planets) have more positive slopes that the Sun.\footnotetext{Note, the [X/Fe]-T$_{\rm c}$ slope values reported by \citet{mel09} must be multiplied by $-1$ before comparing to our values.} We confirmed this finding with our examination of the solar analogs in our sample and the \citet{neves09} sample. 

\citet{mel09} also found that solar analog SWPs have more positive slopes than stars without planets. We confirmed this result in Figures 3 and 5, with an average slope ratio between SWPs and comparison stars ($\sim2.2$) nearly identical to the ratio determined by \citet{mel09}.

In a follow-up study, \citet{ram09} confirmed the findings of \citet{mel09} with additional observations of solar analogs and noted another pattern. They find that the [X/Fe]-T$_{\rm c}$ slope tends to be more negative among the more metal-rich stars. To search for this pattern in our data we have divided the data points from Figure 7 into two groups, [Fe/H] $> 0.10$ and [Fe/H] $\le 0.10$, which we show in Figures 10 and 11, respectively.

\begin{figure}
  \includegraphics[width=3.3in]{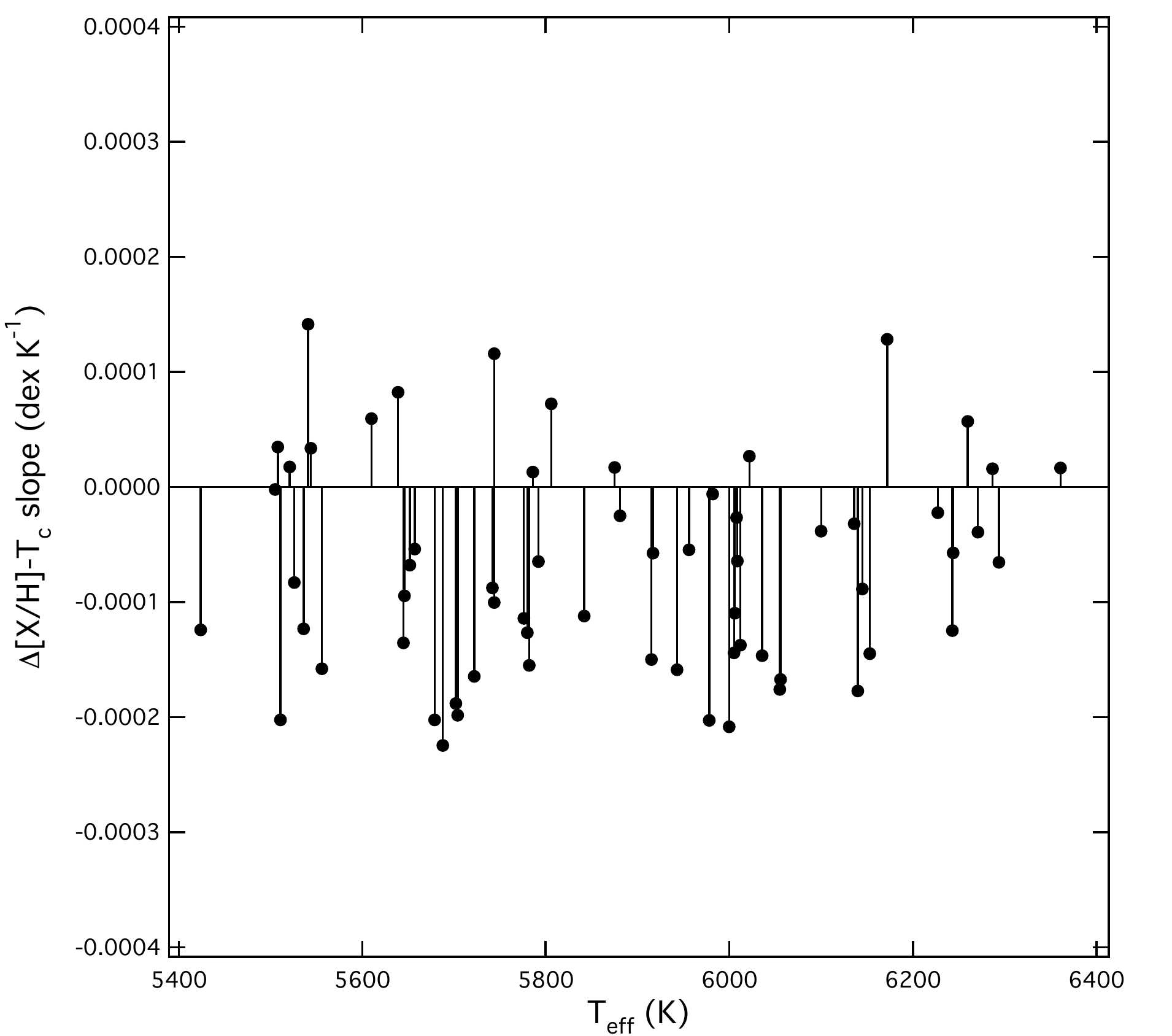}
 \caption{Same as Figure 7, but only showing stars with [Fe/H] $> 0.10$.}
\end{figure}

\begin{figure}
  \includegraphics[width=3.3in]{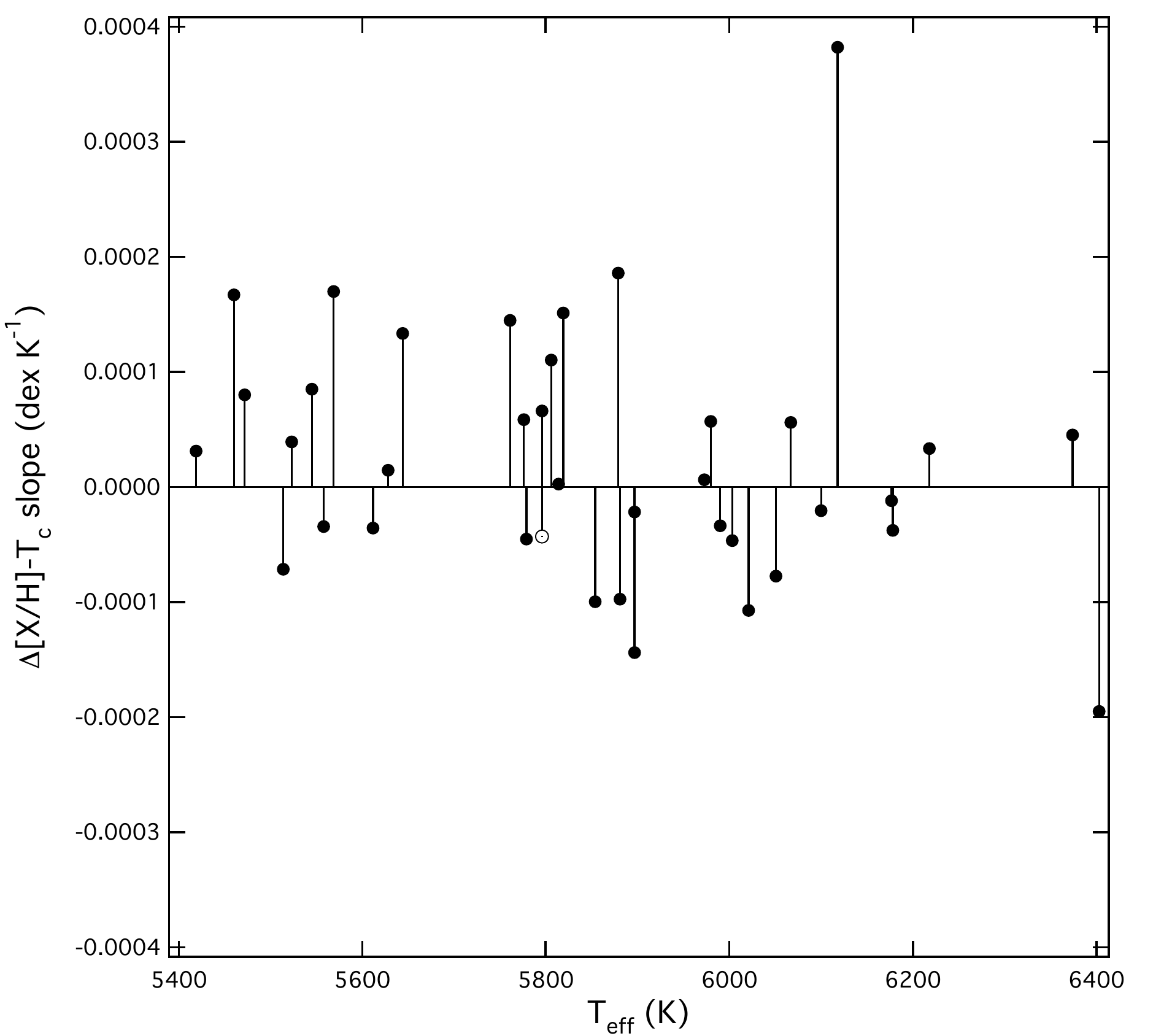}
 \caption{Same as Figure 7, but only showing stars with [Fe/H] $\le 0.10$.}
\end{figure}

The differences between the trends in Figures 10 and 11 are dramatic. In Figure 10 there are 48 SWPs with negative and 15 with positive slopes; restricting the count to values above 0.00007 and below -0.00007 dex K$^{\rm -1}$, there are 32 SWPs with negative and 5 with positive slopes. In Figure 11 there are 17 SWPs with negative slopes and 21 with positive slopes; restricting the count to values above 0.00007 and below -0.00007 dex K$^{\rm -1}$, there are 7 SWPs with negative and 10 with positive slopes.. These results confirm \citet{ram09} but for SWPs over a much more broad temperature range.

From Figure 11, we can see that the Sun has a more negative slope than most SWPs near the solar temperature. Only one SWP in the figure (HIP 116906, one of the solar analogs from \citet{neves09}) has a negative slope (actually very close to the solar value). Several SWPs hotter and a few cooler than the Sun also have negative slopes. Only a larger sample over a similar range of temperatures will show if this is a statistical fluke or a significant anomaly for the Sun.

Our new results for SWPs, combined with our recent results from \citet{gg08} and \citet{gg10} concerning Li abundances and vsini, likely share the same origin. As we suggested in \citet{gg08}, low Li abundance and slow rotation among SWPs likely resulted from the presence of a massive protoplanetary disk around each star; \citet{bv08} has also argued for such a relationship. While modern stellar evolution models that incorporate non-standard mixing are able to reproduce the low Li abundance of the Sun and SWPs (e.g., \citet{do09}), it is still necessary to invoke another parameter to explain why single stars have different rotational histories.

 \citet{ram09} and \citet{mel09} suggest that refractory elements are relatively more depleted in the Sun because they were removed to form terrestrial planets from the gas in the protoplanetary disk, gas which was later accreted onto it. Perhaps a similar process occurs for the SWPs. It is also important to note that hotter dwarfs have less massive envelopes. Therefore, hotter SWPs should be more sensitive to alteration of surface abundances by accretion. There appears to be evidence for this in Figure 7.

The correlation evident in Figures 8 and 9 are relevant to the interpretation of the results of Figure 11. It can help us to determine the source of the Sun's apparently anomalous composition. If refractory elements are relatively more depleted in the Sun because they were removed to form terrestrial planets from gas that was later accreted onto it, then Figures 8 and 9 are telling us that some other process must also be at work. These figures show that the Sun's photosphere is enriched in refractory elements relative to the CI meteorites. If, instead the CI meteorites were depleted in refractory elements when they formed, the responsible process would have had to operate very early in the protoplanetary nebula, given the very primitive nature of these meteorites.

In the case of the Solar System, however, the connection to terrestrial planets should be made using the meteorites formed in the inner Solar System \citep{alex01} and not the primitive CI meteorites that probably formed in the outer Solar System and at earlier times. Since the inner Solar System meteorites are enhanced in refractory elements by about a factor of two relative to the CI meteorites \citep{alex01}, and since we have shown that the Sun is slightly ($\sim$ 0.05 dex) enhanced in refractory elements with respect to the CI meteorites, we conclude that the Sun is deficient in refractory elements (as already suggested by \citet{mel09}). The sequestration of refractory elements into the terrestrial planets was likely an important process.

When discussing elemental abundance trends among stars, it is important to consider the effects of Galactic chemical evolution. Due to the design of our samples and analysis method, however, chemical evolution effects should not be a major influence on our results. This is because we have been careful to compare SWPs to otherwise very similar stars lacking Doppler detectable planets. Nevertheless, it is still possible that there is a mix of thick and thin disk stars in our samples, especially for the metal-poor stars. Thick disk stars have enhanced $\alpha$-element abundances compared to thin disk stars for the same value of [Fe/H]; the possible effects of this difference on planet formation has been discussed in \citet{gg09}. How the different mix of elements in a thick disk star affects the [X/H]-T$_{\rm c}$ slope (e.g., for the data in Figure 11) is beyond the scope of the present work, but it is one that should be addressed as the sample size of metal-poor SWPs increases.

\section{Conclusions}

Using new abundance analyses of SWPs and stars without known planets, we have found that SWPs tend to have more negative [X/H]-T$_{\rm c}$ slopes than stars without planets. Our results confirm \citet{ram09}, who focused their study on solar analogs.

We also find that SWPs with [Fe/H] $> 0.10$ tend to have more negative [X/H]-T$_{\rm c}$ slopes than more metal-poor SWPs, showing that the process responsible for these trends is sensitive to metallicity. 

We revisited the question of the abundances in the solar photosphere relative to CI meteorites and confirmed previous findings that a significant trend with T$_{\rm c}$ exists. The Sun is slightly ($\sim$ 0.05 dex) enhanced in refractory elements relative to the CI meteorites, but compared to the inner Solar System meteorites, the Sun is deficient by almost a factor of two. This implies that sequestration of the refractory elements into terrestrial planets left their marks in the distribution of Solar System abundances.

It appears that both the enrichment of refractory elements in the solar photosphere via accretion and the sequestration of refractory elements into the terrestrial planets left their marks in the distribution of Solar System elemental abundances.

These results, combined with our recent findings that SWPs have lower Li abundances (confirmed by \citet{israel09} for solar analogs) and rotate slower than comparison stars, place new stringent constraints on planet formation models \citep{bv08}. 

\section*{Acknowledgments}

We thank the anonymous referee for helpful comments and suggestions. We also thank Nathaniel Simpson, a student at Grove City College, for assistance with some of the calculations. We acknowledge financial support from the Discovery Institute in Seattle, WA and Grove City College.

\bsp

\label{lastpage}

\end{document}